\documentclass[reprint,showpacs,superscriptaddress,amsmath,amssymb,aps,prb]{revtex4-1}
\usepackage{graphicx}
\usepackage{bm}
\begin{document}

\title{Local Electron Beam Excitation and Substrate Effect on the Plasmonic Response of Single Gold Nanostars}
\author{Pabitra Das}
\affiliation{Surface Physics Division, Saha Institute of Nuclear Physics, 1/AF Bidhannagar, Kolkata 700 064, India}
\author{Abhitosh Kedia}
\affiliation{Department of Physics and Astrophysics, University of Delhi, Delhi 110007, India}
\author{Pandian Senthil Kumar}
\affiliation{Department of Physics and Astrophysics, University of Delhi, Delhi 110007, India}
\author{Nicolas Large}
\affiliation{Department of Electrical and Computer Engineering, Laboratory for Nanophotonics, Rice University, Houston, Texas 77251-1892 }
\author{Tapas Kumar Chini}
\email[Corresponding author : ]{tapask.chini@saha.ac.in}
\affiliation{Surface Physics Division, Saha Institute of Nuclear Physics, 1/AF Bidhannagar, Kolkata 700 064, India}

\begin{abstract}
We performed Cathodoluminescence (CL) spectroscopy and imaging in a high resolution scanning electron microscope to locally and selectively excite and investigate the plasmonic property of a multibranched gold nanostar on silicon substrate. This method allows us to map the local density of optical states from the nanostar with a spatial resolution down to a few nanometers. We resolve both in spatial and spectral domain, different plasmon modes associated with the nanostar. Finite-difference time-domain (FDTD) numerical simulations are performed to support the experimental observations. We investigate the effect of substrate on the plasmonic property of these complex shaped nanostars. The powerful CL-FDTD combination helps us to understand the effect of the substrate on plasmonic response of branched nanoparticles.
\end{abstract}

\maketitle
\section{Introduction}
The localized surface plasmon resonance (LSPR) of noble metal nanoparticles (NPs) can give rise to strong optical scattering and dramatic enhancement of the near-field around the particle.\cite{bohren2008absorption,maier2007plasmonics} LSPR induced local-field enhancement has potential implications in the fundamental understanding of light-matter interaction at the nanoscale and has important application in biomolecular manipulation, labeling, detection \cite{schultz2000single,cao2002nanoparticles} and hyperthermia-based therapy.\cite{bosman2013surface,ye2011plasmonic} The LSPR properties of NPs depend upon several factors, like the size, shape, materials, and environment. Consequently, the ultimate goal towards a controlled synthesis and optical tunability of metal nanostructures has motivated researchers to synthesize and study NPs of different shapes, sizes, and compositions using micro-spectroscopic tools that can provide information at the nanoscale and both in spectral and spatial domain.
In this context, multi-branched Au NPs are an exciting new class of structures having interesting physicochemical properties. These unique multi-branched NPs have enormous applications in catalysis,\cite{lim2011metal} plasmonics,\cite{guerrero2011nanostars,hao2007plasmon} sensing,\cite{dondapati2010label} as well as in surface enhanced Raman scattering (SERS)\cite{rodri않uez2009surface,hrelescu2009single} and are thought of as the building blocks of nanoscale devices. The multi-branched NP (also called multi-pod, star-shaped NP, or nanostar) consists of a core region with several tips protruding in all directions. Their LSPRs are tunable into the near infrared region\cite{rodri않uez2009surface} and are of interest for biomedical applications.\cite{bosman2013surface, ye2011plasmonic} Optical properties of these nanostars have been investigated for the last few years by means of UV-VIS spectroscopy,\cite{hao2004synthesis,kumar2008high, kedia2012precursor} dark-field microscopy (DFM),\cite{shao2012plasmonic} SERS,\cite{rodri않uez2009surface,hrelescu2009single} and photoelectron emission microscopy (PEEM),\cite{hrelescu2009single} as well as by numerical simulation methods such as finite-difference time-domain (FDTD) simulations,\cite{hao2007plasmon,shao2012plasmonic} finite element method (FEM),\cite{yuan2012gold} discrete dipole approximation (DDA),\cite{aaron2008polarization} and boundary element method (BEM).\cite{kumar2008high, mazzucco2011spatially} Optical techniques like UV-VIS spectroscopy probe ensembles of NPs and are consequently limited by large inhomogeneous broadening induced by the size and shape distribution of the ensemble. Even single particle spectroscopy technique like DFM can probe a metal nanoparticle as a whole and therefore cannot distinguish the rich features arising out from the different parts of the highly asymmetric structures as in nanostar. Another interesting and promising technique to probe localized surface plasmons at the single NP level is scanning near-field optical microscopy (SNOM).\cite{bakker2007near} However, this technique requires the fabrication of very delicate probe in the form of a nanoscale sharp tip and very often suffers from the drawback of the probe-sample interaction that distorts the signal of interest. An excellent alternative is electron microscopy based techniques, i.e. electron energy loss spectroscopy (EELS),\cite{bosman2013surface, mazzucco2011spatially, mayoral2012nanoscale, koh2009electron} and cathodoluminescence (CL) spectroscopy.\cite{kuttge2009local,vesseur2007direct,chaturvedi2009imaging,gomez2008mapping,denisyuk2010transmitting} CL in a transmission/ scanning electron microscope (TEM/SEM) through the detection of the emitted photons, or EELS in TEM through the detection of energy loss suffered by the inelastically scattered transmitted electrons are shown to constitute an excellent probe of plasmons that allow few nanometer resolution information in the spatial domain.\cite{das2012probing, das2012spectroscopy, myroshnychenko2012plasmon, knight2012aluminum} The excitation mechanism of LSP modes with an electron beam is quite different from that of the optical excitation. An electron or a beam of electrons moving in vacuum can act as an evanescent source of super continuum light.\cite{de2010optical} The CL-SEM is based on the fact that energy is coupled from electrons to the plasmon modes of the metallic nanostructure and subsequently to the propagating light modes that constitute one of the prominent decay channels for plasmons. While a light source, as employed in DFM, excites the whole volume of the nanoparticle, a finely focused electron beam can act as a local probe to excite plasmons and consequently a CL image gives access to the electromagnetic local density of states (LDOS). Although EELS has been applied to probe the plasmon modes of isolated multi-branched metallic nanoparticles with very high spatial and spectral resolution in some earlier studies,\cite{bosman2013surface, mazzucco2011spatially} CL has not yet been applied for such geometry to the best of our knowledge. Moreover, EELS suffers from the drawback of samples to be electron transparent. On the other hand CL in SEM do not have this limitation and can be efficiently applied to obtain correlated single particle spectroscopic and microscopic information including the effect of the substrate which is not easily probed by EELS. Here, by means of CL, we locally and selectively probe LSP modes of an isolated Au nanostar on Si substrate, and acquire the emission pattern from different location of the same NP. CL imaging and spectra are compared with 3D-FDTD simulations for a simple model nanostructure composed by a spherical core and a single branch. Numerical simulations were performed to assess the substrate effect on the LSPRs as a function of the tip-substrate separation, probing the symmetry breaking induced by the substrate.\cite{knight2009substrates, zhang2011substrate, wu2009finite} Very often, for the sake of simplicity or computational resource limitations, the substrate along with the medium above the substrate is modeled as an effective medium surrounding the nanostructure. As a result, very rich plasmonic features arising from the reduced symmetry and breaking of degeneracy are not described. Consequently, detailed analyses of the scattering spectra of the nanostar geometry that includes the effect of substrate is of utmost importance for theoretical understanding of plasmonic response from nanometer length scale complex-shaped metallic objects sitting on a dielectric substrate. This is also important in the optimization of near-field enhancement for the particles exhibiting spikes/tips and used in SERS.\cite{rodri않uez2009zeptomol} Application of site-selective CL spectroscopy and imaging augmented with systematic numerical analysis gives us new physical insight into the complex nature of the local plasmonic properties of multi-branched nanostructures on a substrate.

\section{Experimental}
\subsection{Sample Preparation }
A solution of Polyvinylpyrrolidone, PVP (MW=10000, Aldrich) in DMF (Merck) is homogeneously mixed with adequate amount of aqueous solution of tetrachloroauric acid (HAuCl$_4$.3H$_2$O, Aldrich) so as to make the molar ratio of PVP to metal as ~3250 (calculated in terms of polymer repeating unit or monomer chain length)\cite{kedia2012precursor}. The whole solution is then continuously stirred under normal ambient conditions. The color of the solution starts changing from pale yellow to colorless and then finally becomes blue, indicating the stable formation of gold nanostars. The as-formed Au nanostars were centrifuged seven fold at 4500 rpm and repeatedly washed with triple distilled water so as to remove the excess free PVP. The final sample is carefully drop-coated on silicon substrates and directly used for CL imaging after dried up.
\subsection{CL Spectroscopy and Imaging}
CL and electron beam induced radiation emission (EIRE) imaging on an isolated single Au nanostar on Si substrate were performed in a ZEISS SUPRA40 SEM equipped with the Gatan MonoCL3 cathodoluminescence system. The ZEISS SUPRA40 SEM has a hot Schottky field emission gun (FEG) and the attached MonoCL3 system uses a retractable paraboloidal light collection mirror. The parabolic mirror collects light that is emitted from the sample covering 1.42$\pi$ sr of the full 2$\pi$ of the upper half sphere and collimates it through a hollow aluminum tube to a 300 mm Czerny-Turner type optical monochromator and finally the signal is fed to a high-sensitivity photomultiplier tube (HSPMT). Data were obtained with an electron acceleration voltage of 30 kV and beam current of about 15 nA with a beam diameter of $\sim$5 nm. The electron beam was directed onto the sample surface through a 1 mm diameter opening in the mirror. To ensure maximum efficiency of light collection, the top surface of the sample is kept at the focal plane of the mirror, which lies approximately 1 mm below the bottom plane of the mirror. Before every set of experiment we adjusted this optical focal plane with utmost care using the stepper motor controlled sample stage. Schematic of the CL setup is shown in Figure 1c.
The CL system in conjunction with the SEM can be operated in two modes, namely, monochromatic and panchromatic. In monochromatic mode, the focused e-beam is either scanned over the sample or positioned on a desired spot. The emitted light from the sample passing through the monochromator allows the emission spectra to be recorded serially. To minimize sample drift during acquisition, we have acquired the experimental spectra in two wavelength ranges separately: 500-700 nm and 700-900 nm, respectively, with a step of 4 nm. The total acquisition time of one spectrum was approximately 12.5 s. Spectra have been averaged for each e-beam position and corrected from the substrate background. The monochromatic photon map is then built up at a selected peak wavelength of the EIRE spectrum by scanning the e-beam over the sample. 
For each e-beam position, the luminescence is collected over the entire sample. The bright pixels then correspond to the areas where the strongly excited plasmon mode emits the photons. When adding all the position dependent partial maps, obtained for each e-beam position, we obtain a full CL map of the plasmon mode associated to a particular wavelength. This CL image is proportional to the radiative local density of optical states (LDOS) of the plasmonic structure.\cite{kuttge2009local, vesseur2007direct}
\subsection{FDTD simulations for electron beam excitation}
To further understand surface plasmon assisted photon emission from the Au nanostars considered here, we have performed 3D-FDTD numerical simulations (Lumerical Solutions). Maxwell's equations are solved in discretized space and discretized time to follow the response of a material to any applied EM field (i.e. the evanescent field associated with e-beam in case of CL). The current density associated to the e-beam is given by, 
\begin{equation}
J(t,\vec{r}) = -ev\hat{u_z} \delta(z-vt) \delta(x-x_0) \delta(y-y_0) , 
\end{equation}
where {\it e} is the electronic charge, and {\it v} is the velocity of electron, $(x_0, y_0)$ represents the position of the electron beam and $z$ is the direction of electron velocity and $\hat{u_z}$ is the unit vector along {\it z} direction. This current density can be modeled as a series of electric dipoles with temporal phase delay ($z/v$) (here $v$ = 0.32$c$ corresponding to the 30 keV electron energy used in the present experiment). We used the experimental dielectric permittivity tabulated by Johnson and Christy for gold,\cite{johnson1972optical} and a refractive index of 4 for silicon.
\section{Results and Discussion}
As seen in the secondary electron (SE) image there is slight polydispersity in the size distribution of the nanostars (Supporting Info. Figure S1). Observations under SEM also reveal that most of the Au nanostars are aggregated and form clusters. However, one isolated single Au nanostar was selected for the CL study. From SE image (inset in Figure 1a) the particle has four major sharp tips (marked as A, B, C, and D) lying approximately in the same horizontal plane (i.e., orthogonal to the incident e-beam). Other tips are protruding out of this plane. The overall morphology is flower-like with crisscrossed non-planar single crystalline tips branching out from the complex core.\cite{kedia2012precursor} In the following we will focus on tip A that has a length of 50 nm and an aperture angle of approximately 15$^o$. Figure 1a shows CL spectra taken from different location of tip A of the nanostar (colored dots in the inset of Figure 1a). Spectra clearly show a strong dependence of the CL intensity on the e-beam position. 
\begin{figure}[t]
\centering
  \includegraphics[width=8cm]{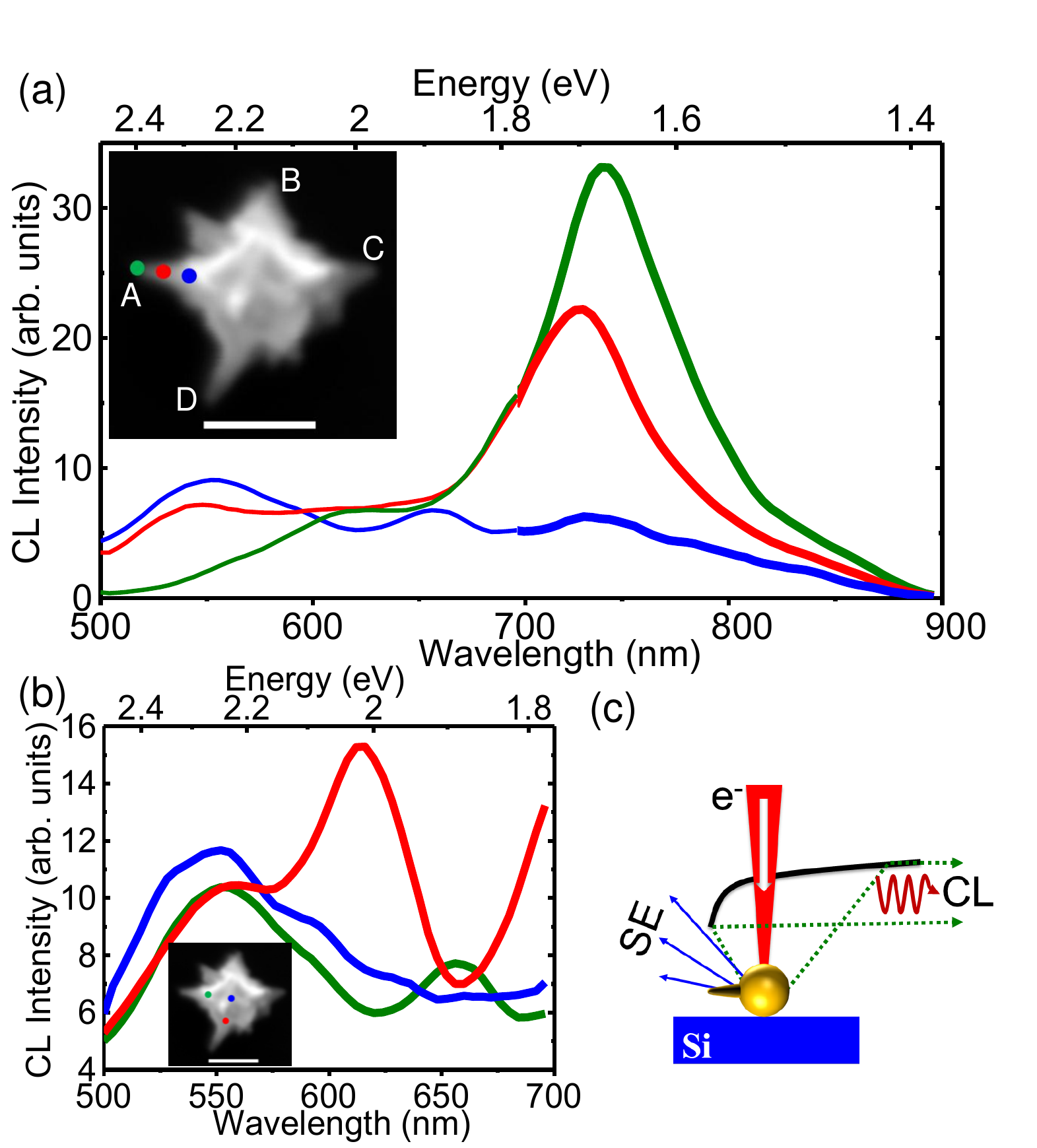}
  \caption{(Color Online)(a) Experimental CL spectra of the Au nanostar. The spectra were acquired in two wavelength ranges separately as 500-700 nm (thin line) and 700-900 nm (bold line) and then merged with proper normalization.  Associated SEM image is shown as inset. (b) CL spectra in the range 500-700 nm from different e-beam positions in the core region. Scale bars correspond to 100 nm. (c) Schematic showing CL experimental setup. }
  \label{fgr:Figure1}
\end{figure}
Two major peaks are observed from CL measurements. The low energy peak is around 750 nm while the high energy one is located between 550 nm and 620 nm (Figure 1a, b). Additionally, another resonance appears around 660 nm when exciting the nanostar at the base of the tip (blue curve in Figure 1a). Interestingly, as we go from the apex region of the tip towards the core region, the low energy peak (750 nm) intensity decreases and the high energy peak (550 nm) intensity increase. To get a better understanding of the effect of the core of the nanostar on the CL response we also acquired CL spectra from three different e-beam positions in the core region (Figure 1b). The CL spectra, corresponding to all the three impact points of the core (inset), exhibit the peak at 550 nm. While the CL from the central portion of the core region is dominated by the 550 nm peak only, we also see resonances at 620 nm and 660 nm for the e-beam impact near the periphery of the core region, i.e. at the tip base.
\begin{figure}[t]
\centering
 \includegraphics[width=8cm]{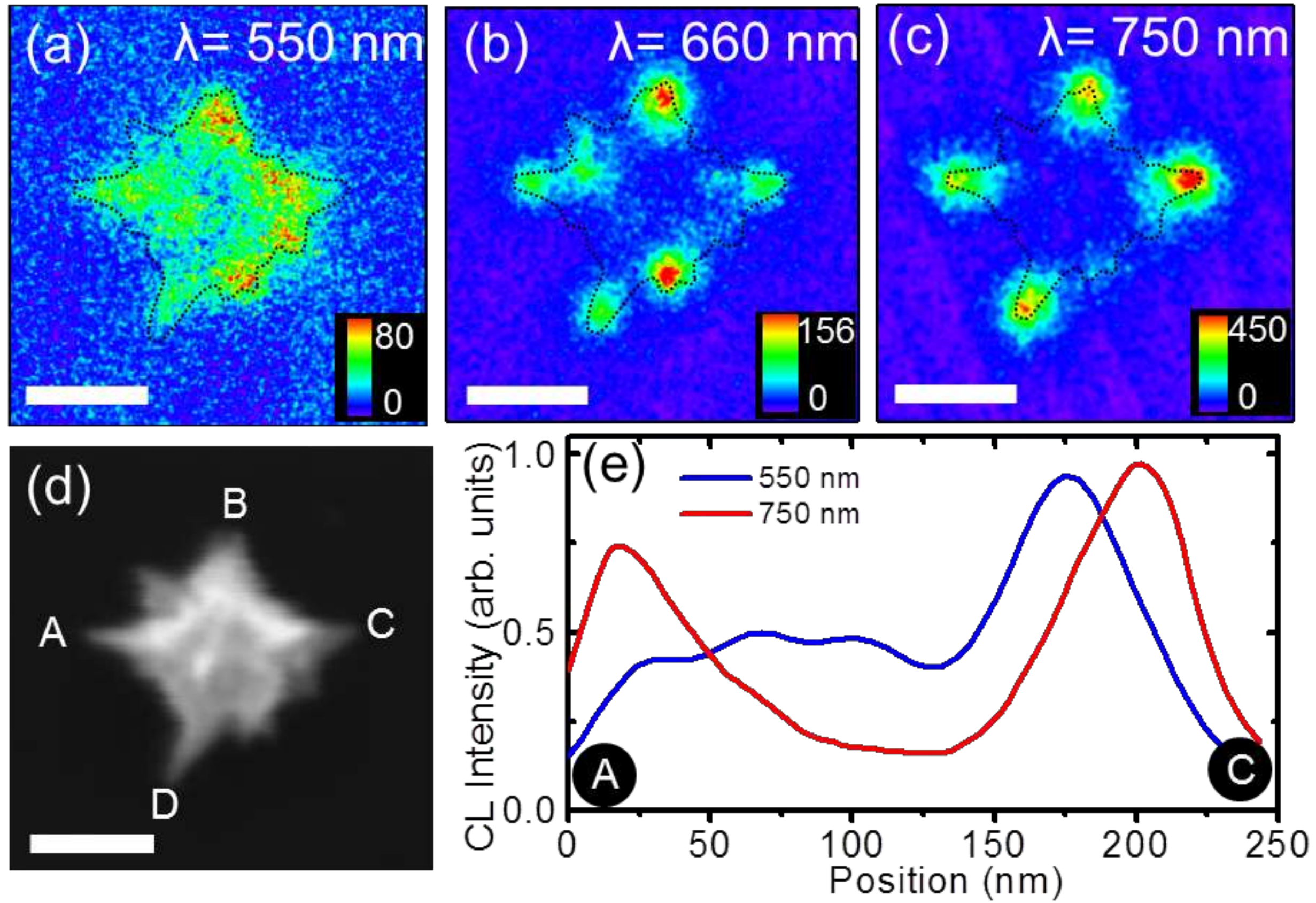}
\caption{(Color Online)Experimental CL imaging on the isolated Au nanostar recorded at (a) 550 nm, (b) 660 nm, and (c) 750 nm. (d) SE image showing the particle morphology. Scale bars correspond to 100 nm. (e) Monochromatic CL profile along the AC path with a 2.5 nm step at 550 nm and 750 nm.}
  \label{fgr:Figure2}
\end{figure}
We acquired monochromatic CL maps at the peak wavelengths (550 nm, 660 nm, and 750 nm) to extract the spatial distribution of the photon emission (Figure 2). The monochromatic CL image obtained at 550 nm wavelength shows luminescent intensity when the electron beam scans over the core region (Figure 2a), whereas the strong light emission occurs at 660 nm (Figure 2b) and 750 nm (Figure 2c) when the e-beam scans over the peripheral region of spherical core and tip apex. The bright spots in the photon maps arise when the LSP modes are resonantly excited by the evanescent field associated with the electron beam. The LSPR at 550 nm is well reported in previous works and is associated with the spherical core region.\cite{hao2007plasmon,shao2012plasmonic} When going towards higher wavelengths (750 nm), the luminescence enhancement is shifted gradually towards the tip region and is strongly confined to the far ends of the longer tips (Figure 2c). This is consistent with the spectroscopic data of Figure 1a. The LSPR at ~660 nm gives rise to photon emission coming from the region where the tip meets the periphery of core (Figures 1 and 2b).\\
We will henceforth designate the resonant peak at 550 nm as the core LSP mode and the 750 nm peak as the tip LSP mode. The tip LSPR positions and intensities depend on the e-beam excitation position on the tip region as well as on the tip dimension. To determine the spatial extent of the tip and core modes we acquired monochromatic CL line profile at 750 nm and at 550 nm by placing the 30 kV electron beam at 100 points along the AC path (2.5 nm step). Results are shown in Figure 2e. At 750 nm, luminescence is seen to be peaked near the two ends giving clear evidence of strong photon confinement of the tip mode to the apex region of the tips (here A and C). Interestingly, comparing the CL maps (Figure 2b, c), one can see that the low energy photons (750 nm) are more susceptible to be emitted from the long tip apex whereas the high energy photons (660 nm) are emitted from the shorter tips or at the base of the longer tips. Moreover, the CL spectroscopic and  imaging data of Figures 1 and 2 reveal that the intensity level of the tip mode is more than five times higher than that of the core mode with the associated LSPR being narrower than that of the core mode (Figure 1a, b). The linescan at 550 nm gives a non-uniform variation of CL intensity pattern directly related to the irregular shape of the core region, induced by the presence of the numerous shorter tips.\\
To better understand the spectral and spatial distribution of the CL emission from the Au nanostar, we calculated the CL spectra. In order to mimic the 1.42$\pi$ solid angle collection path, we used a 10 $\mu$m $\times$  10 $\mu$m 2D power monitor placed 300 nm above the substrate. We assumed our model nanostar structure to be 110 nm spherical core with a single 50 nm long tip. The tip is assumed to be parallel to the Si substrate and has an aperture angle of $\theta$=15$^o$, consistent with the morphology of the nanostar extracted from the SEM images (Figure 2d). The major dipolar LSPR that occurs in this type of structure is above 700 nm.\cite{kumar2008high} We already showed experimentally that tip LSP mode remains near the apex region of the tip and do not penetrate into the core of the NP (Figure 2c). Considering the typical skin depth of gold to be about 15 nm at optical frequencies, it is impossible for the electromagnetic field to pass through the whole nanoparticle.\cite{mazzucco2011spatially} Moreover, the nature of local excitation source in the form of finely focused e-beam with a spot size of $\sim$ 5 nm, as used in the present experiment, decreases the probable interference of other tips in the spectra. Hence the charge oscillation of one tip cannot couple efficiently to the charge oscillation of other tips. The coupling to the core and to other tips is then negligible. Moreover, it has already been reported, from simulations with optical excitations, that the introduction of extra tips to the core structure gives rise to a change of the intensity ratio of the LSPRs. Further increasing the number of tips, with the same dimensions, leads to a very small spectral shift of the resonance (Supporting Info. Figure S2). This justifies the validity of the assumption of single tip and spherical core model for our nanostar.\\
To investigate the effect of the substrate on which the nanostar is sitting calculations have been done with and without Si substrate for three impact parameters (inset in Figure 3). Calculated spectra are dominated by a LSPR at $\sim$750 nm, in agreement with experimental results. In this configuration, the position of this tip mode is almost insensitive to the presence of the substrate. Our simulations also indicate that the presence of the substrate changes only the intensity of the luminescence. We observe a threefold decrease in the tip mode intensity due to the presence of the substrate.
\begin{figure}[t]
\centering
  \includegraphics[width=8cm]{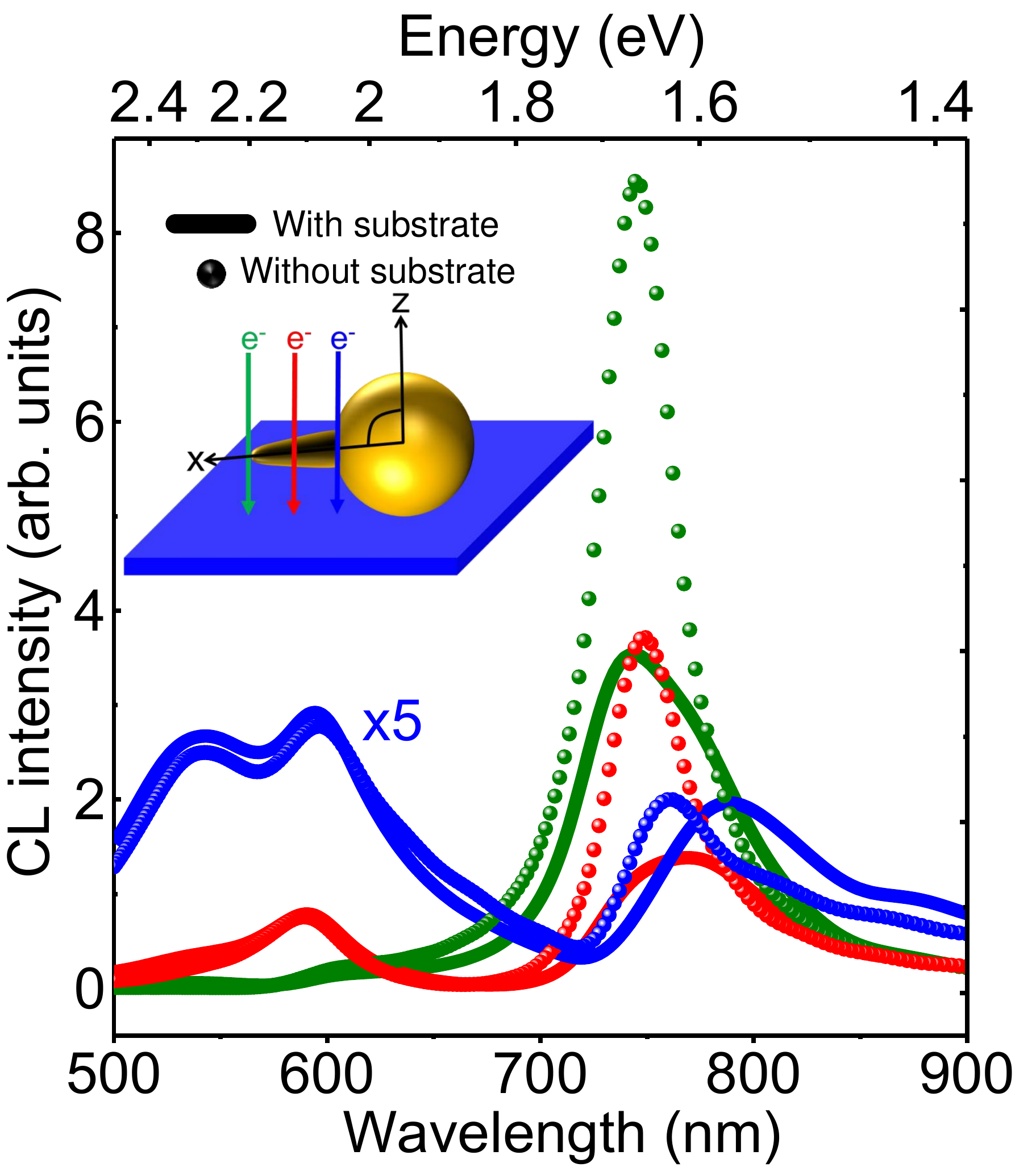}
  \caption{(Color Online)FDTD calculated CL spectra from the model nanostructure (110 nm core, 50 nm horizontal tip with a 15o aperture). Spectra, calculated with the Si substrate (solid lines) and without the substrate (dotted lines), are color-encoded with the e-beam position (marked with colored arrows on the schematics). Spectra corresponding to the tip base excitation (blue spectra) have been magnified by a factor of 5.}
  \label{fgr:Figure3}
\end{figure}
Multi-branched structures are usually formed by tips of different shapes, sizes, and orientations. Relative position of the tips with respect to the substrate plays a very crucial role in the plasmonic properties of the nanostar. To investigate further this effect we vary the angle between the tip and the substrate ($\theta$), maintaining the e-beam excitation on the tip apex. Figure 4 shows CL spectra calculated for different angular tip configurations from parallel ($\theta$=0$^o$) to orthogonal ($\theta$=90$^o$) to the substrate. By rotating the tip in the counter clockwise direction ($\theta$$<$0$^o$), the tip apex approaches the surface (inset in Figure 4a). 
\begin{figure}[t]
\centering
  \includegraphics[width=8cm]{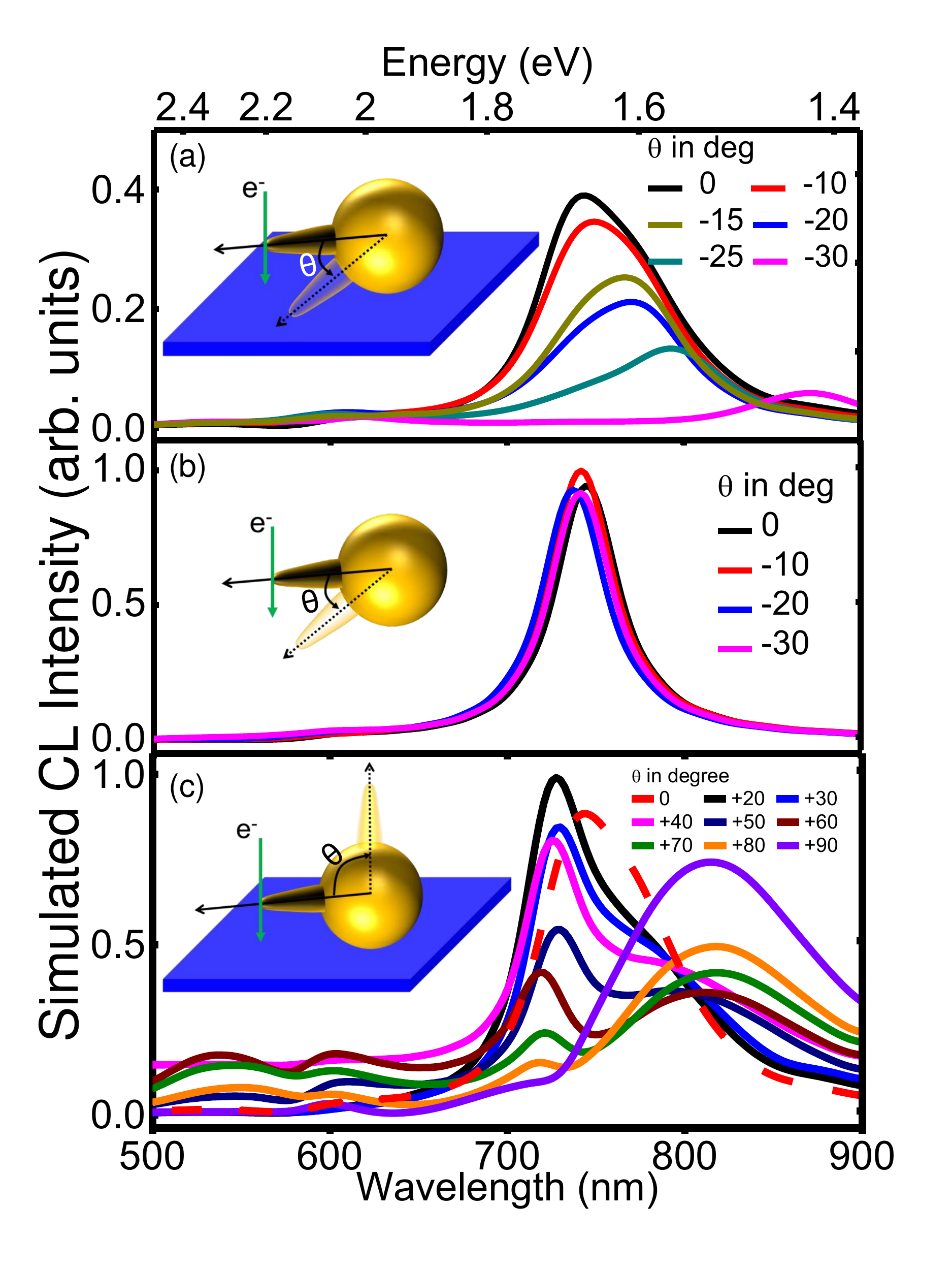}
  \caption{(Color Online)FDTD calculated CL spectra for the model single tip structure as a function of the tip orientation. Spectra are calculated for both the NP (a) with and (b) without the substrate as function of the tip orientation (-30$^o$ $\leq$ $\theta$ $\leq$ 0$^o$). (c) CL spectra calculated for tip moving away from the substrate (0$^o$ $\leq$ $\theta$ $\leq$ 90$^o$).}
  \label{fgr:Figure4}
\end{figure}
The results for -30$^o$ $\leq$ $\theta$ $\leq$ 0$^o$ are shown in Figure 4a, b in presence and in absence of substrate, respectively. When the NP is on top of the substrate the intensity of the tip LSPR decreases and red-shifts (Figure 4a), while there is no noticeable change without the substrate (Figure 4b). Supporting info. Figure S3 shows spectra for higher values of $\theta$. The physical mechanism behind the intensity decrease is preferential scattering of light at the resonance wavelength into the high refractive index substrate which has a higher density of states.\cite{mertz2000radiative, spinelli2011controlling} Both direct near-field absorption in the substrate and radiative emission into the substrate introduce a loss channel for the light confined in the particle, thus yielding broadening (i.e. damping) of the resonances in the scattering spectrum. Now let us consider the case when the tip is rotated in the clockwise direction ($\theta$$>$0$^o$), i.e. tip apex going away from the substrate (Figure 4c). With increasing $\theta$, the plasmon mode splits at two wavelengths one blue-shifted and the other one red-shifted with respect to the original peak at 750 nm for $\theta$= 0$^o$ (dashed line). For angles over 50$^o$ a distinct peak appears at 830 nm which prevails. This peak is associated to the dipolar LSPR of the spherical core (Supporting Info. Figure S4). As we increase the angle $\theta$, and maintain the excitation on the tip apex, the electron beam effectively comes closer to the core. When the tip is in out-of-plane configuration ($\theta$= 90$^o$) and remains at the farthest distant from the substrate, the modes of the spherical core are efficiently excited. In such configuration, the CL spectrum is dominated by the LSP mode of the spherical core supported by the silicon substrate. With increasing $\theta$ the observed blue-shift of the tip LSPR is associated with the uncoupling the tip and the substrate (screened by the NP core) It has already been reported that higher order LSPRs can be activated by the presence of the substrate.\cite{knight2009substrates,zhang2011substrate} LSPRs experimentally observed around 550 nm when exciting the core region may be associated to the quadrupolar LSPR of the spherical core (Supporting Info. Figure S4). From a direct comparison of the CL map (Figure 2a) to the SEM image (Figure 2b) we can notice that the maximum intensity regions of the CL map matches the bright regions of the SEM image. This indicates that the associated LSPR comes from much smaller tips pointing upward. As simulations indicate that a higher order core mode is also be located around 550 nm, the CL signal experimentally observed at the wavelength is then a mixture of short tip mode and higher order core mode. Drastic spectral changes, induced by the position of the tips with respect to the substrate, are critical for surface enhanced spectroscopies and sensing applications. However, to the best of our knowledge, this aspect had not been discussed so far in literature for nanoparticles exhibiting spikes or tips.\\
\begin{figure}[t]
\centering
  \includegraphics[width=8cm]{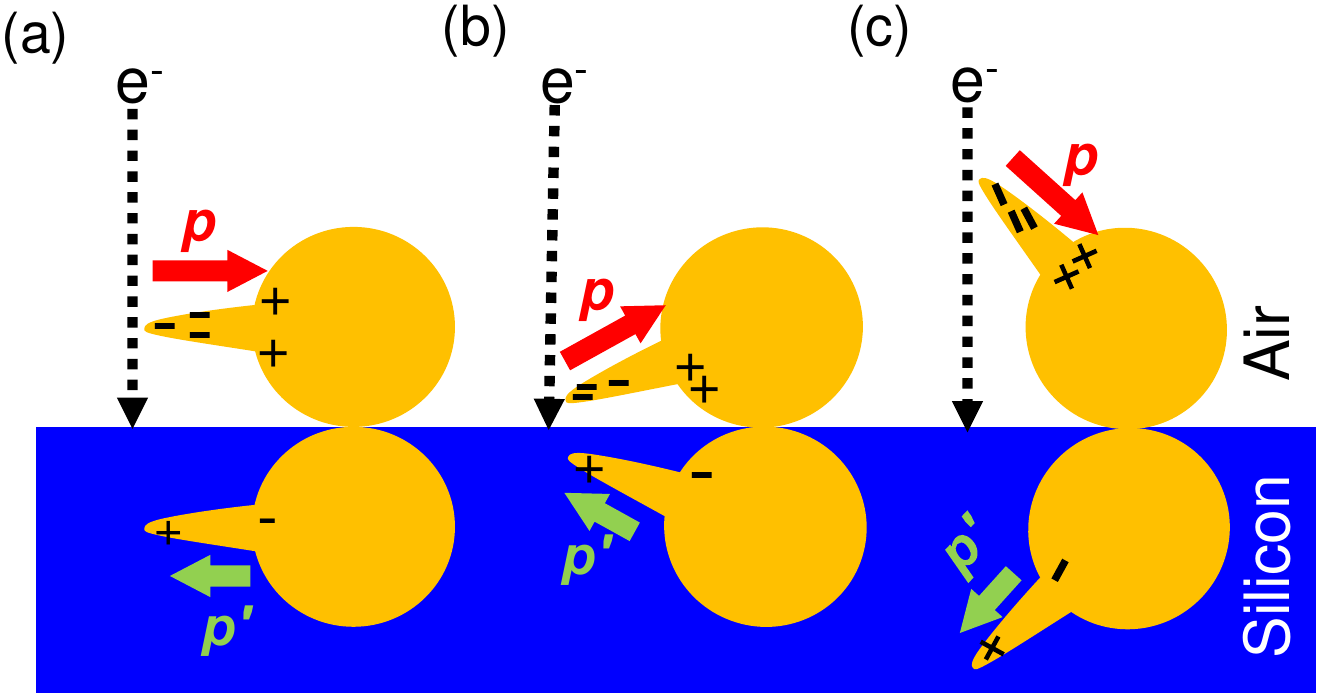}
  \caption{(Color Online)Schematic of the image charge distributions induced by the substrate for (a) a tip parallel to the substrate ( $\theta$= 0$^o$), and (b) a tip pointing toward the substrate ($\theta$ $<$ 0$^o$) and (c) a tip pointing away from the substrate ($\theta$ $>$ 0$^o$). The magnitude of the image charge depends upon the dielectric permittivity of the substrate. The NP dipole moment $p$ and its image $p'$ are in the opposite direction. Relative amplitudes of the dipole moments are given by the arrow sizes.}
  \label{fgr:Figure5}
\end{figure}
The red-shift of the LSPR in the presence of a substrate can be better understood with the help of an image charge model that incorporates the symmetry breaking induced by the substrate. A spherical NP embedded in a homogeneous medium has three degenerate dipolar LSP modes. Introduction of a dielectric substrate beneath the NP reduces the symmetry and lifts the degeneracy, giving rise to a splitting of the dipolar LSPR.\cite{knight2009substrates} In such case, the interaction between the nanoparticle and the dielectric substrate can be viewed as the NP interacting with its own image inside the substrate.
The image charge magnitude is reduced by a factor ($\epsilon$-1)/($\epsilon$+1), where $\epsilon$ is the dielectric permittivity of the substrate. The strength of particle-image interaction is governed by the separation between the nanoparticle and the substrate, the substrate permittivity, and the polarization of the incident radiation. Using electron beam as the excitation source has some particular importance in this regard. Indeed, an electron beam has an evanescent electric field aligned with the propagation direction of the beam, and a magnetic field along the direction perpendicular to the e-beam. So electron beam can efficiently excite both the modes along the beam and perpendicular to the beam. Schematics of charge and image charge distributions for our model single branched nanostructure on silicon substrate are shown for $\theta$= 0$^o$, $\theta$ $<$ 0$^o$, and $\theta$ $>$ 0$^o$ (Figure 5a, b, and c, respectively). At $\theta$= 0$^o$ (Figure 5a) the image charge of the tip has a slightly weaker dipole moment $p'$, and pointing in the opposite direction, than the actual NP dipole moment $p$ reducing the net dipole moment of the entire system (reduction of scattering intensity observed in Figure 4a, b). More specifically the oscillation of the induced image charges is out of phase with that excited in the particle itself.  When the tip approaches the substrate surface (Figure 5b), charges move closer leading to a stronger NP-image interaction (red-shift, Figure 4a). On the other hand, when the tip moves away from the substrate (Figure 5c), charges are localized further from the substrate surface, giving rise to a blue-shift of the LSPR (Figure 4c).\\
We have already shown that the introduction of the substrate for the $\theta$= 0$^o$ configuration (tip parallel to the substrate) does not alter the peak position, but only reduces the LSPR intensity. Hence, for the sake of simplicity and numerical efficiency, we further neglect the substrate and assume an isolated nanostructure.
 \begin{figure}[t]
\centering
  \includegraphics[width=8cm]{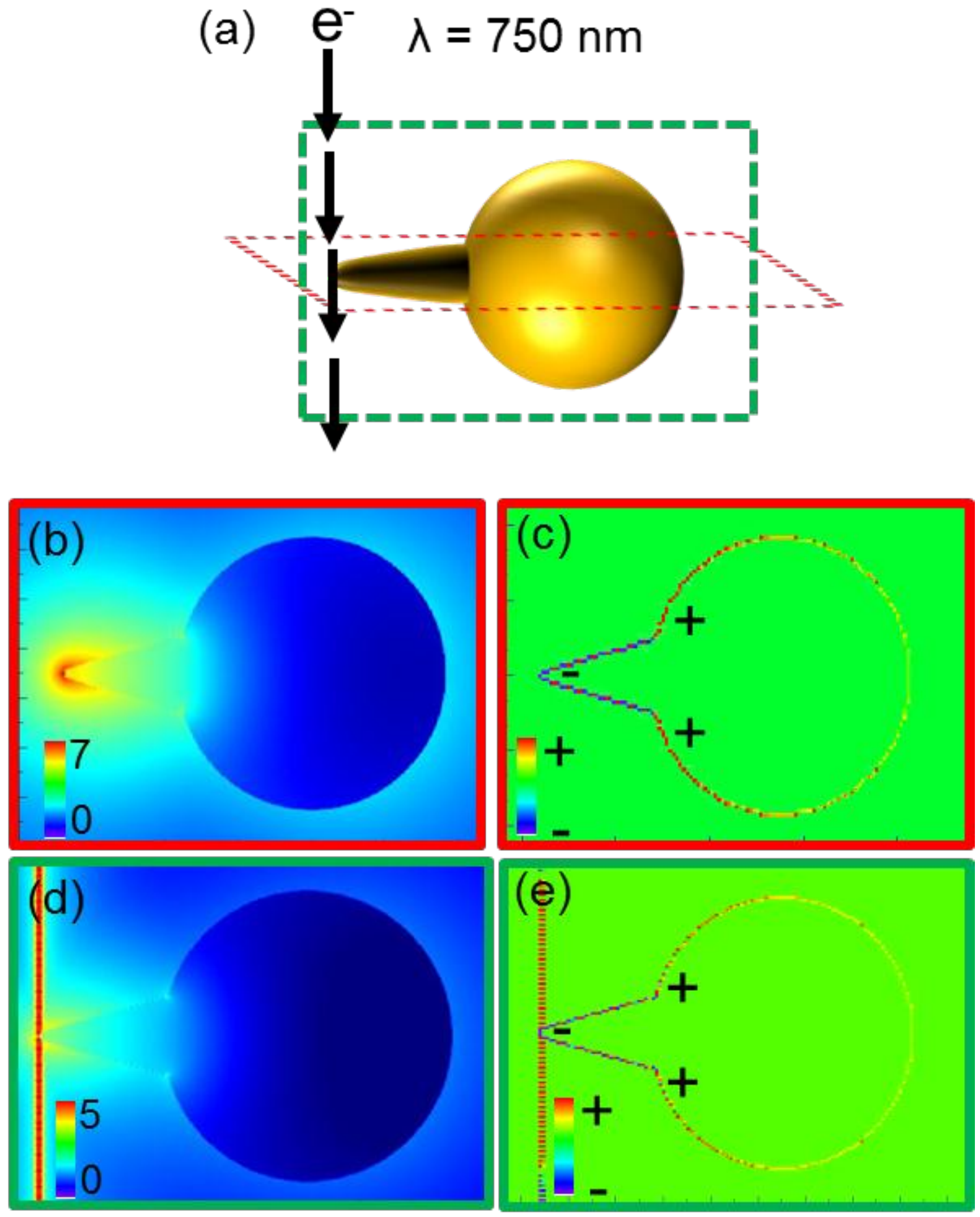}
  \caption{(Color Online)Local electric field intensity distribution and surface charge distribution calculated at 750 nm from the plane orthogonal to the e-beam (b and c, resp.) and the plane containing the e-beam (d and e, resp.). (a) Schematic showing the cross section planes with respect to the e-beam and to the nanoparticle.}
  \label{fgr:Figure6}
\end{figure}
To confirm whether the 750 nm LSPR is of dipolar nature, we calculated the near-field intensity distribution (Figure 6b, d) and the associated surface charge distribution (Figure 6c, e). Results are shown for the plane orthogonal to the e-beam and the plane containing the e-beam (cf. Figure 6a). The charge distribution clearly demonstrate that the 750 nm resonance peak is due to a dipolar oscillation along the length of the tip and is more efficiently excited when the e-beam is focused near the tip apex.
Here, we have assumed a perfect sphere supporting a single tip, therefore presenting only a single contact point with the substrate. However, the actual nanostar exhibits multiple tips pointing out in various directions, thus leading to the possibility of multiple contact points with the substrate. Such geometrical consideration drastically complicates the interpretation of the nanostar-substrate interaction. Furthermore, other structural effects such as roughness of the spherical core (caused by enormous number of smaller tips present in the nanostar),\cite{oubre2004optical} roundness of the tips,\cite{large2011plasmonic} and substrate oxidation\cite{davies2012metal} can also modify the spectral properties of the nanostar. Apart from these approximations calculations are in good agreement with the experimental results.
\section{Conclusion}
CL spectroscopy and imaging technique was used for an isolated gold nanostar supported on a silicon wafer to demonstrate the spectrally and spatially resolved photon emission from the tip and core LSP modes of a single gold nanostar. Good agreement was found between experimental results and numerical simulations. We have shown that local morphology, associated to substrate effects, leads to spectral changes in the plasmonic response of the nanostar. Indeed, the relative position of the nanostar tips with respect to the substrate strongly modified the tip-substrate interaction, thus leading to spectral shift of the tip LSPR.  Furthermore, the relative tip-core LSPR contributions have been investigated as function of the tip angular configuration. We have shown that spectra of nanostars where the tips point toward the e-beam are dominated by the core LSPR. Our experiment and simulation results demonstrate that the long tips in multibranched nanostars can serve as efficient emitters at optical frequency that can be precisely tuned by engineering the tip-substrate separation. Local plasmonic properties, due to the morphology, combined with substrate effects have important consequences for a wide range of applications in surface-enhanced spectroscopies, sensing, ultrasensitive optical analyses, design of optical devices, and non-linear photonics requiring a methodology for tackling the substrate effect. We believe, that high resolution CL technique combined with detailed numerical approaches as used here can open a novel route towards the analysis of the fine plasmonic behavior of substrate supported complex morphology nanoparticles.
\begin{acknowledgements}
The authors wish to thank P. Nordlander for his insight and input.
\end{acknowledgements}

\bibliography{ref}
\end{document}